\documentstyle[12pt,amstex]{article}

\newcommand{\na}[1]{\nabla_{#1}}
\def\tf{\tilde{f} }
\def\tR{\tilde{R} }
\def\RR{ I \hspace*{-0.8ex} R }
\def\intS{\int\limits_{S_\epsilon}\hspace*{-0.8ex} dA }
\def\intV{\int\limits_{{\RR}^3 - B_\epsilon}\hspace*{-0.8ex}d^3 x }
\def\tha{\vartheta(a -\rho) }
\def\sqa{\sqrt{a^2-\rho^2} }
\def\da{\delta(\rho- a) }
\def\CM{\cal M}

\begin{document}
\renewcommand{\thefootnote}{\fnsymbol{footnote}}
\newpage
\pagestyle{empty}
\begin{center}
{\LARGE {
Distributional Energy-Momentum Tensor \\ \medskip
of the   Kerr-Newman Space-Time Family
}}\\
\vfill
\vfill

\vspace{2cm}
{\large
Herbert BALASIN
\footnote[7]{ e-mail: hbalasin @@ email.tuwien.ac.at}
\footnote{partially supported by the ``Hochschuljubil\"aumsstiftung d. Stadt
Wien`` project H-00013 }
}\\
{\em
Institut f\"ur Theoretische Physik, Technische Universit\"at Wien\\
Wiedner Hauptstra{\ss}e 8--10, A - 1040 Wien, AUSTRIA
}\\[.5cm]
{\em and}\\[.5cm]
{\large Herbert NACHBAGAUER
\footnote[8]{e-mail: herby @@ lapphp1.in2p3.fr}
}\\
{\em{Laboratoire de Physique Th\'eorique}}
{\small E}N{\large S}{\Large L}{\large A}P{\small P}
\footnote{URA 14-36 du CNRS, associ\'ee \`a l'E.N.S. de Lyon,
et au L.A.P.P. (IN2P3-CNRS)\\
\hspace*{0.7cm} d'Annecy-le-Vieux}
\\
{\em Chemin de Bellevue, BP 110, F - 74941 Annecy-le-Vieux Cedex,
France}\\[.5cm]
\end{center}
\vfill
\begin{abstract}
Using the Kerr-Schild decomposition of the
metric tensor that employs the algebraically special nature of the
Kerr-Newman space-time family, we calculate the energy-momentum tensor.
The latter turns out to be a well-defined tensor-distribution with
disk-like support.\\

\end{abstract}
\vfill

\rightline{{\small E}N{\large S}{\Large L}{\large A}P{\small P}-A-453/93}
\rightline{TUW 93 -- 28 }
\rightline{gr-qc/9312028}
\newpage

\renewcommand{\thefootnote}{\arabic{footnote}}
\setcounter{footnote}{0}
\newpage
\pagebreak
\pagenumbering{arabic}
\pagestyle{plain}

\section*{\Large   Introduction}
Singular space-times present one of the major challenges in general
relativity. Originally it was believed that their singular nature
is due to the high degree of symmetry of the well-known
examples ranging from the Schwarzschild geometry to the
Friedmann-Robertson-Walker
cosmological models.  However, Penrose and Hawking \cite{HaEl} have shown in
their
celebrated singularity theorems that singularities are a phenomenon
which is inherent to general relativity.
Since the standard approach allows only for smooth space-time
metrics, one has to exclude the so-called singular regions from the space-time
manifold. \par
In a recent work \cite{BaNa} the authors advocated the
use of such  distributional techniques to calculate the energy-momentum
tensor of the Schwarzschild geometry. It turns out that it is
possible to include the singular region (i.e. the space-like line
$r=0$ with respect to  Schwarzschild coordinates) in the space-time which
now no longer is a vacuum geometry, and to identify it with the
support of the energy-momentum tensor. The latter becomes
a tensor-distribution \cite{Li,Pa} with delta-like shape.\par
This reasoning puts the Schwarzschild geometry on the same footing with its
ultrarelativistic limit the Aichelburg-Sexl shock-wave geometry \cite{AiSe},
where the energy-momentum tensor has a delta-like support
on a null line and is interpreted as being generated by a
particle moving with velocity of light. Adopting this line of
arguments one might consider
the Schwarzschild geometry as being generated by a tachyon
which hides behind the event horizon thus providing a new
interpretation of cosmic censorship.\par
The aim of the present work is to extend our calculation to the general
axisymmetric, stationary space-time family discovered by Kerr and
Newman \cite{KeNe,BoLi}. This family also contains
the Schwarzschild geometry and
its charged extension the Reissner-Nordstr{\o}m solution as special cases
of spherical symmetry. We will show
that the distributional techniques developed in
\cite{BaNa} apply
to this family too thereby allowing to calculate
its energy-momentum  tensor.

\noindent
One of the main features of the Kerr-geometry (and its charged
version) which eventually led to its discovery is  the existence of a
geodetic, null vector-field $k^a$ which gives rise to a corresponding
congruence of geodesics. Taking advantage of this fact it is possible to
decompose the metric into the so-called Kerr-Schild form \cite{KeSc},
$ g_{ab} = \eta_{ab} + fk_a k_b, $
where $\eta_{ab}$ denotes a flat (background) metric and $f$ a scalar field.
This geometrical decomposition greatly facilitates the calculation not only
from  a technical point of view but also from a conceptual one since it
identifies $k^a$ as being an integral part of the geometry which will
be kept fixed during an eventual regularisation of $g_{ab}$.\par
Our work is organised in the following way: The first chapter
is devoted to the calculation of the Ricci-tensor and the
curvature scalar of an arbitrary metric of Kerr-Schild form.
In the second chapter we will rederive our previous results concerning the
Schwarzschild geometry and extend them to the Reissner-Nordstr{\o}m case.
Finally, in the third section we calculate the distributional
energy-momentum tensor  of  the Kerr- and Kerr-Newman geometries.

\section*{\Large  1) Algebraically special geometries and Kerr-Schild
structure}
All geometries we are going to consider in this work are algebraically
special geometries, which allow a Kerr-Schild decomposition of the metric
\begin{equation}\label{KS1}
g_{ab}=\eta_{ab} + f k_a k_b,
\end{equation}
where $k^a=\eta^{ab}k_b$ denotes a null vector (field) with respect to the
metric $\eta _{ab}$, which in turn implies its nullity with
respect to $g_{ab}$.   The above
decomposition provides two metrical structures associated with a given
manifold $\CM$, $g_{ab}$ and $\eta_{ab}$ respectively. In what follows
$\na{a}$ denotes the derivative operator associated with $g_{ab}$ and
$\partial_a$ its commuting counterpart with
respect to $\eta_{ab}$.
An important consequence of the nullity of $k^a$ and the decomposition
(\ref{KS1}) is
\begin{equation}\label{GEO}
k^a\na{a} k^b = k^a\partial_a k^b,
\end{equation}
which can be derived from the explicit form of the difference operation
$C^a{}_{bc}$ of the derivative operator
\begin{align}\label{CHR}
&\na{a} v^b = \partial_a v^b + C^b{}_{ad}v^d , \qquad v^a \in
\Gamma(T\CM ),\\
&C^a{}_{bc} = \frac{1}{2}g^{ad}\left(\partial_b g_{dc} +
\partial_c g_{db} -
\partial_d g_{bc}\right)\nonumber\\
&= \frac{1}{2}\left(\partial_b(fk^ak_c) +
\partial_c(fk^ak_b)-\partial^a(fk_bk_c)
+ f k^a (k\partial)( f k_bk_c )  \right),\nonumber
\end{align}
where all index-raising and lowering is done with  respect to $\eta_{ab}$.
Equation (\ref{GEO}) tells us that if $k^a$ is geodetic with respect to
$g_{ab}$  the same is true with respect to $\eta_{ab}$ and vice-versa.
Kerr and Schild \cite{KeSc}
have shown that the vacuum Einstein-(Maxwell)-field equations
require $k^a$ to be geodetic, and we  can assume $k^a$ to be affinely
parametrised,  i.e. $( k\partial )  k^b=0$.
This condition is an essential property of the geometry and thus
we will  strictly maintain it, even in the course of a regularisation
procedure.\par
The calculation of the Ricci-tensor using the conventions \cite{Wa}
\begin{align*}
&R_{abd}{}^c=\partial_bC^c{}_{ad} - \partial_aC^c{}_{bd} + C^c{}_{fb}C^f{}_{ad}
- C^c{}_{fa}C^f{}_{bd},\\
&R_{ab} = R_{acb}{}^c = \partial_cC^c{}_{ab} - \partial_aC^c{}_{cb} +
C^c{}_{cf}C^f{}_{ab}- C^c{}_{af}C^f{}_{cb},\
\end{align*}
together with the form (\ref{CHR}) of $C^a{}_{bc}$,
which implies $C^a{}_{ab}=0$,  yields
\begin{align}\label{KSRIC}
&R^a{}_b = \frac{1}{2}\left (\partial^a\partial_c(fk^ck_b) +
\partial_b\partial^c(fk_ck^a)  -\partial^2(fk^ak_b) \right ),\\
&R= \partial_a\partial_b(fk^ak^b)\nonumber.
\end{align}
Let us again remind the reader that from now on all indices are raised
and lowered with the help of $\eta_{ab}$. (\ref{KSRIC}) has the remarkable
property of being a sum of second derivatives linear with respect to
$f$ which is neither the case for the upper nor the lower index parts.
This property will allow a distributional evaluation of (\ref{KSRIC})
whenever $fk^ak^b$ is itself a well-defined distribution and therefore
possesses a natural second derivation.

\section*{\Large  2) Schwarzschild geometry and Reissner Nordstr{\o}m
extension}
A simple illustration of the above concepts is provided by the Schwarzschild
geometry and its Reissner-Nordstr{\o}m extension.
The Kerr-Schild form of the Schwarzschild geometry, whose line
element reads in Schwarzschild coordinates
$$
ds^2 = -dt^2\left ( 1-\frac{2m}{r}\right ) + dr^2
\left (1-\frac{2m}{r}\right )^{-1} + r^2 d\Omega^2,
$$
is most easily displayed using the coordinate transformation
$$
\bar{t} = t - 2m\log (2m -r), \qquad r < 2m
$$
which  gives
\begin{equation}\label{SSKN}
ds^2 =-d\bar{t}^2+dr^2+ r^2d\Omega^2+ \frac{2m}{r}(d\bar{t}-dr)^2
\end{equation}
and  allows the immediate identification
$$
f=\frac{2m}{r},\quad k^a=(1,e^i_r),\qquad i=1,2,3
$$
where $e^i_r$ is the unit vector with respect to $\eta_{ab}$ of the
spherical-coordinate system. Due to the stationarity of the metric (\ref{SSKN})
the distributional evaluation of the Ricci-tensor and the curvature-scalar
reduce to a 3-dimensional problem on the $\bar{t}=const$ surfaces.
Let us evaluate the curvature scalar explicitly acting on
an arbitrary test-function $\varphi \in\>C^\infty_0(\RR ^3)$,
\begin{align}
&\label{5a}
(R,\varphi) = \lim_{\epsilon\to 0}\>\intV \>f k^i k^j\>\partial_i
\partial_j \varphi(x)=  \\
&= \lim_{\epsilon\to 0} \left[
- \intS \>f k^i k^j\>N_i \partial_j \varphi  (\epsilon  e^i_r) +
\intS \>\partial_i (fk^i k^j ) N_j \varphi (\epsilon  e^i_r)
\> + \right. \nonumber \\
& \hspace*{5cm}
\left. + \intV \> \varphi (x) \partial_i \partial_j ( f k^i k^j ) \right]
\nonumber
\end{align}
where $B_\epsilon$ is the $\epsilon$-ball around the origin and
$S_\epsilon$ its boundary. $N^i = r^2 \sin \theta\> e^i_r$ is the outward
directed normal of $S_\epsilon$ and $dA$ denotes $d\theta d\phi$.
The last term in (\ref{5a}) is just the curvature scalar which vanishes in the
region   $\RR^3 - B_\epsilon $, the first term
is already of order $\epsilon$. Thus only the second term contributes
and we get in the limit $\epsilon \to 0$
$$
(R,\varphi) = \lim_{\epsilon\to 0 } \> \intS   \frac{2m}{\epsilon^2}\>
\epsilon^2\sin  \theta\>\varphi (\epsilon  e^i_r)
=  8\pi m\> \varphi (0)
$$
which is precisely the result we obtained in \cite{BaNa} in Schwarzschild
coordinates.
With respect to Kerr-Schild  coordinates the energy-momentum tensor becomes
$$
8\pi\>T^a{}_b=R^a{}_b - \frac{1}{2}\delta^a{}_b R = -8\pi m\>\delta^{(3)}(x)
(\partial_{\bar{t}})^a\>(d\bar{t})_b.
$$
The extension of this result to the Reissner-Nordstr{\o}m metric is easily
achieved by observing that we only have to replace
$$
f=\frac{2m}{r} \quad\rightarrow\quad f+\tf = \frac{2m}{r} - \frac{e^2}{r^2}.
$$
Since expression (\ref{KSRIC}) is linear in  $f$ we necessarily
obtain an additional contribution $\tR^a{}_b$ to the Ricci-tensor.
Note that $\tilde f $ is still locally integrable, which is a
necessary condition for the existence of the distribution $R$.
Let us exemplify the calculation of $\tR^a{}_b$
by the evaluation of $\tR^0{}_0$.
\begin{align*}
&(\tR^0{}_0,\varphi) = \lim_{\epsilon\to 0}
\frac{1}{2}
\intV\>\tf\partial^2
\varphi =\\
& \frac{1}{2}     \lim_{\epsilon\to 0}  \left[
-  \intS\> \tf (N\partial)\varphi + \intS\>(N\partial)\tf\>\varphi +
\intV\>\partial^2\tf\>\varphi \right]=\\
& \frac{1}{2}
\left[  \intS \left(\frac{e^2}{\epsilon^2}\right )\epsilon^2\sin
\theta(e_r\partial)\varphi (0)  + \intS\>\epsilon^2\sin\theta
\left(\frac{2e^2}{\epsilon^3}\right )(\varphi(0) + \epsilon
(e_r\partial)\varphi(0))\right. \\
& \hspace*{2cm}
\left. - \intV \left(\frac{2e^2}{r^4}\right )
(\varphi(x)  -\varphi(0)) -
\varphi(0) \intV\> \left(\frac{2e^2}{r^4}\right)\right] =\\
&-e^2\int d^3 x\>\frac{1}{r^4}(\varphi(x)-\varphi(0)) =:
-(\left[\frac{e^2}{r^4}\right ],\varphi),
\end{align*}
where beginning with the third line of the calculation the
limit $\epsilon\to 0$ is considered implicitly thereby dropping all terms
of order $\epsilon$. The calculation of the remaining components proceeds
along the same lines so that  we get in the end
$$
\tR^0{}_0 = -\left[\frac{e^2}{r^4}\right ],
\quad \tR^0{}_i = 0,
\quad \tR^i{}_j = \left[ \frac{e^2}{r^4}(\delta^{ij} - 2e^i_r e^j_r)\right],
\quad \tR =0 .
$$
The electromagnetic part of the energy-momentum
tensor is still traceless which tells us that the regularisation
procedure did not destroy the conformal invariance of the theory.\par
With respect to Kerr-Schild coordinates
the total energy-momentum tensor of the Reissner-Nordstr{\o}m geometry
is given by
$$
T^0{}_0 = -m\delta^{(3)}(x) - \frac{e^2}{8\pi}\left[\frac{1}{r^4}\right
], \quad
T^0{}_i=0,\quad
T^i{}_j = \frac{e^2}{8\pi}\left[ \frac{1}{r^4}(\delta^{ij} - 2e^i_r
e^j_r)\right].
$$
This result emphasizes the fact that the electromagnetic part of $T^a{}_b$
is a well-defined tensor-distribution, which coincides with the classical
result for $r\neq 0$, i.e.
$$
(\left[\frac{e^2}{r^4}\right ],\varphi) = (\frac{e^2}{r^4},\varphi)
$$
for all $\varphi\in\>C^\infty_0(\RR ^3)$ that vanish at $r=0$.

\section*{\Large 3) Kerr and Kerr-Newman geometries}
Passing from Schwarzschild to Kerr generalises the null
vector-field $k^a$ and the function $f$ to
\begin{align*}
&k^a = (1,k^i) = (1,\frac{rx+ay}{r^2 + a^2},\frac{ry-ax}{r^2 + a^2} ,
\frac{z}{r})\\
&f=\frac{2mr}{\Sigma},\qquad\Sigma = \frac{r^4+a^2z^2}{r^2},
\end{align*}
where $r$ is implicitly given by
\begin{equation}\label{CONS}
r^4 - r^2 (x^2 + y^2 + z^2 -a^2) -a^2z^2 =0.
\end{equation}
The cartesian coordinate system $\{x,y,z\}$ refers to the background metric
$\eta_{ab}$ of the Kerr-Schild decomposition. Taking again advantage
of the stationarity of the metric reduces the distributional evaluation to
a 3-dimensional problem. The constraint (\ref{CONS}) may be solved by a
change of coordinates from cartesian to spheroidal
\begin{align*}
&x=\sqrt{r^2+a^2}\sin\theta\cos\phi\\
&y=\sqrt{r^2+a^2}\sin\theta\sin\phi\\
&z=r\cos\theta.
\end{align*}
These coordinates represent a deformation
of spherical coordinates expressed by the parameter $a$. The $r=const$
surfaces become confocal ellipsoids
with focus on the ring $\rho^2 =x^2+y^2=a^2,\>z=0$, whereas the $\theta=const$
surfaces are hyperboloids with an asymptotic cone of aperture $\theta$.
For $r=0$ the ellipsoid degenerates into a double cover of the disk
$\rho \leq a,\>z=0$. In the latter region the coordinates essentially
reduce to polar coordinates $(\rho,\phi )$ with radius $\rho = a \sin \theta $.
\par
The calculation proceeds in a fairly straightforward fashion, using the
general formulas of chapter one. This time, however, we have to exclude a
disk-shaped region $r\leq\epsilon$ from the integrals and consider the
limit $\epsilon\to 0$ afterwards in order to do partial integrations.
To illustrate the procedure explicitly let us calculate the curvature-scalar
given by (\ref{5a})
where $N_i$ denotes the surface normal of the $r=const$ surface and $dA$
the coordinate-surface area-element. Taking into account the identities
\begin{align}
&N^i = r\sqrt{r^2 + a^2}\sin^2 \theta e^i_\rho + (r^2+a^2)\sin\theta
\cos\theta e^i_z, \label{IDKERR}  \\
&k^i = \frac{\sin\theta}{\sqrt{r^2+a^2}}(re^i_\rho - ae^i_\phi) +
\cos\theta e^i_z, \quad (Nk) = \Sigma \sin \theta  , \nonumber\\
&(N\partial)k^i = \frac{a\sin^2\theta}{\sqrt{r^2+a^2}}
( a e^i_\rho + r e^i_\phi ) ,
\nonumber\\
&(k\partial)f + f (\partial k) =\frac{2m}{\Sigma},\nonumber
\end{align}
where $e^i_\rho$, $e^i_\phi$ and $e^i_z$ denote the unit vectors of
the cylindrical coordinates of $\RR ^3$, facilitates the
evaluation of (\ref{5a})
\begin{align*}
&(R,\varphi) = \intS \frac{2m}{\Sigma} \>\Sigma\sin\theta
\>\varphi(a\sin\theta\cos\phi ,a\sin\theta\sin\phi ,0) =\\
&\frac{4m}{a} \int\limits^{2\pi}_0 \! d\phi\int\limits^a_0
\frac{\rho d\rho}{\sqa}
\varphi(\rho\cos\phi ,\rho\sin\phi ,0)
\end{align*}
which finally gives
$$
R(x) = \frac{4m}{a}\frac{\tha }{\sqa} \delta (z).
$$
The calculation of the components of the Ricci-tensor proceeds along similar
lines and gives
\begin{align}\label{RICKERR}
&R^0{}_0 = 2m \left[ \frac{a\tha }{\sqa ^3}\right ]
\delta(z) - \frac{2m}{a}\da\delta(z),\\
&R^0{}_i = 2m\left[ \frac{\rho\tha }{\sqa ^3}\right ]
\delta(z)e^i_\phi - \frac{m\pi}{a}\da\delta(z)e^i_\phi,\nonumber\\
&R^i{}_j = -\frac{2m}{a}\left[\frac{\rho^2\tha}{\sqa ^3}
\right ]\delta(z)e^i_\phi e^j_\phi +
\frac{2m}{a}\frac{\tha}{\sqa}\delta(z) e^i_z e^j_z + \nonumber \\
& \hspace*{8cm} \frac{4m}{a} \da\delta(z) e^i_\phi e^j_\phi
\nonumber
\end{align}
where
\begin{align*}
&(\left[ \frac{\tha}{\sqa ^3}\right ] \delta(z) ,\varphi) := \\
&\int\limits^a_0 d\rho\int\limits^{2\pi}_0 d\phi
\frac{\rho}{\sqa ^3}\left( \varphi(\rho\cos\phi,\rho\sin\phi,0)-
\varphi(a\cos\phi,a\sin\phi,0) \right)
\end{align*}
defines the distribution denoted by the square brackets. Our result shows
that the energy distribution is concentrated on a disk with radius $a$ in
the \mbox{$z=0$}
plane. For all test functions that vanish on the circle $\rho=a$
our result
coincides with that obtained in the  classical work \cite{Is}
on the source of the Kerr-geometry. Moreover, it can be
shown that  the limit $a\to 0$
exists and coincides with the result obtained in the Schwarzschild case.
Let us demonstrate this fact for the curvature scalar
\begin{align*}
&\lim_{a\to 0}(R_{K},\varphi)= \lim_{a\to 0}
4m\int\limits^a_0d \rho\int\limits^{2\pi}_0
d\phi \frac{\rho}{a\sqa }\varphi(\rho e^i_\rho,0) =\\
&8\pi m \int\limits^1_0 \frac{xdx}{\sqrt{1-x^2}}\>\varphi(0) = 8\pi m
\varphi(0)= (R_S,\varphi),
\end{align*}
where $R_K$ and $R_S$ denote the curvature scalars of the Kerr and the
Schwarzschild geometry respectively. \par
Passing from Kerr to its electromagnetic extension the Kerr-Newman geometry
merely changes the function
$$
f=\frac{2mr}{\Sigma}\>\to\>f+\tf = \frac{2mr}{\Sigma} - \frac{e^2}{\Sigma}
$$
in the Kerr-Schild decomposition (\ref{KS1}), as it was the
case with Schwarzschild and Reissner-Nordstr{\o}m. The innocent looking
additional contributions due to $\tf$ turn out to  require considerable
computational effort , although all integrals that are involved can be
done analytically. In order to spare the reader the  unwieldy formulas
we will
just give  some useful identities needed for the calculation
\begin{align}\label{KNID}
&\partial_i k_j = \frac{r}{\Sigma} T_{ij} -
\frac{a\cos\theta}{\Sigma}\epsilon_{ijk} k^k, \quad
T_{ij} = \delta_{ij} - k_i k_j , \\
&(k\partial)\tf + (\partial k)\tf = 0 ,\quad
(N\partial)\tf = \frac{2e^2(r^2+a^2)r\sin\theta}{\Sigma^2}\nonumber
\end{align}
where
$\epsilon_{ijk}$ the standard $\epsilon$-tensor of $\RR ^3$. Together
with the identities (\ref{IDKERR}), (\ref{KNID}) leads to the following
result for the $\tf$-contribution $\tR^a{}_b$ to the Ricci-tensor
and to the curvature scalar
\begin{align}\label{KNRIC}
\tR^0{}_0 &= -\frac{3\pi e^2}{4} \partial_i\left (
\frac{\da}{a}\delta(z)e^i_\rho \right )-
\left [ \frac{e^2(r^2+a^2+a^2\sin^2\theta)}{\Sigma^3} \right ],\\
&\nonumber\\
\tR^0{}_i &= -\frac{e^2}{a}\partial_k\left
(\frac{\tha}{\sqa}\delta(z)(e^k_\rho e^i_\rho +
e^k_\phi e^i_\phi)\right) - \frac{3\pi e^2}{4}\partial_k
\left(\frac{\da}{a}\delta(z)e^k_\rho e^i_\phi\right )\nonumber\\
&- \frac{\pi e^2}{2a}\left( \frac{\da}{a}\delta(z)(e^i_\rho -
e^i_\phi)\right ) + e^2\left[\frac{\rho}{a}  \frac{\tha}{\sqa^3}\delta(z)
e^i_\rho\right ] \nonumber\\
&-2e^2\left[\frac{a\sin\theta}{\Sigma^3}\sqrt{r^2+a^2}e^i_\phi\right ],
\nonumber\\
&\nonumber\\
\tR^i{}_j&=\frac{2e^2}{a}\partial_k\left(
\frac{\rho}{a}   \frac{\tha}{\sqa}
\delta(z)(e^k_\rho e^{(j}_\rho e^{i)}_\phi + e^k_\phi
e^{(j}_\phi e^{i)}_\phi \right)\nonumber\\
&+\frac{3\pi e^2}{4} \partial_k\left(
\frac{\da}{a}\delta(z)e^i_\phi e^j_\phi e^k_\rho \right )
-\frac{2e^2}{a^2}\left[ \frac{\rho^2\tha}{\sqa^3}\delta(z) \right ]e^{(i}_\rho
e^{j)}_\phi\nonumber\\
&+ \frac{4e^2}{a}\frac{\da}{a}\delta(z)e^{(i}_\rho
e^{j)}_\phi
+\left[ \frac{e^2}{\Sigma^3}((r^2+a^2)\cos^2\theta-r^2\sin^2\theta)
(e^i_\rho e^j_\rho - e^i_z e^j_z) \right ]\nonumber\\
&+\left[ \frac{e^2}{\Sigma^3}(r^2+a^2 + a^2\sin^2\theta) e^i_\phi
e^j_\phi\right ]
- \frac{4e^2}{\Sigma^3}r\sqrt{r^2+a^2}\sin\theta\cos\theta e^{(i}_\rho
e^{j)}_z,\nonumber\\
&\nonumber\\
\tR &= \frac{2e^2}{a} \partial_k \left ( \frac{\rho}{a}   \frac{\tha}{\sqa}
\delta(z) e^k_\phi \right ),\nonumber
\end{align}
where the parenthesis around the indices of the base vectors denotes
symmetrisation with unit weight. The non-vanishing trace of the
Ricci-tensor is due to the length-scale $a$ of the disk.
However, conformal invariance is restored in the limit $a \to 0$,
thus reproducing the result of the spherical symmetric case.
\par
Let us finally comment on the issue of regularisation. Our calculation
started from the Kerr-Schild decomposition of the metric. In the derivation
of the Ricci-tensor and the curvature-scalar we implicitly assumed the
validity of classical differential calculus and the existence of
a smooth function
$f$. This might be  interpreted as regularising the
intermediate quantities like the difference tensor of the covariant
derivatives. Since the final result, the Ricci-tensor, turned out to
be the second derivation of a distribution it was possible to directly
evaluate it without referring to any regularisation procedure.

\section*{\Large Conclusion}
In the present work we have explicitly shown how to calculate the
distributional energy-momentum tensor of the Kerr-Newman space-time
family, thereby following closely the approach proposed in \cite{BaNa}
which includes the singular regions of the geometry in the manifold.
The Kerr-Schild-structure related to the algebraically special nature
of this space-time family turned out  to play a prominent role of the
interpretation. Moreover, our results furnish a well-defined basis
for the investigation of the so-called ultrarelativistic limit geometries,
\cite{AiSe,LoSa} by boosting the energy-momentum tensor. Work in this direction
is currently under progress.
\vfill
\noindent{\em Acknowledgement:\/} The authors are greatly indebted
to Prof.~P.~C.~Aichelburg for many useful discussions.

\vfill

\newpage
\begin{thebibliography}{99}
\bibitem{HaEl} Hawking S W and Ellis G F R, {\em The large scale structure
of space-time}
Cambridge University Press, 1973.\\
\bibitem{BaNa} Balasin H and Nachbagauer H, {\em Class.~Quantum Grav.~}
{\bf 10}, 2271 (1993).\\
\bibitem{Li} Lichnerowicz A, {\em Propagateurs,
Commutateurs et Anticommutateurs en Relativit\'e G\'en\'erale},
IHES No.~10 (1961).\\
\bibitem{Pa}  Parker P E, Distributional Geometry,
{\em J.~Math.~Phys.~}{\bf 20} (1979) 1423.
\bibitem{AiSe}Aichelburg P and Sexl R,
{\em J.~Gen.~Rel.~Grav.~}{\bf 2} (1971) 303.\\
\bibitem{KeNe} Kerr R P, {\em Phys.~Rev.~Lett.~}{\bf 11} (1963) 237;\\
Newman E T and Janis A I, {\em J.~Math.~Phys.~}{\bf 6} (1965) 915.
\bibitem{BoLi} Boyer R H and Lindquist R W,
{\em J.~Math.~Phys.~}{\bf 8} (1967) 265.
\bibitem{KeSc} Debney G C, Kerr R P and Schild A,
{\em J.~Math.~Phys.~}{\bf 10} (1969) 183.\\
\bibitem{Wa} Wald R, {\em General Relativity}, University of
Chicago Press, 1984.\\
\bibitem{Is} Israel W, {\em Phys.~Rev.~}{\bf D 2} (1970) 641;
        Hamity V H, {\em Phys.~Lett.~}{\bf A56} (1976) 77.
\bibitem{LoSa} Loust\'o C O and S\'anchez N,
{\em Nucl.~Phys.~}{\bf B383} (1992) 377.\\
\end{thebibliography}
\end{document}